\documentclass[letter]{aa} 

\usepackage{graphicx}
\usepackage{url}
\usepackage{natbib}
\bibpunct{(}{)}{;}{a}{}{,} 
\usepackage{color}

\def\79com{X0338}
\def\XMM{XMM-\textit{Newton}\ }
\begin{document}

   \title{Discovery of the X-ray selected galaxy cluster XMMU\,J0338.8+0021 at $z=1.49$\thanks{Based on observations under programme ID 084.A-0844 collected at the European Organisation for Astronomical Research in the Southern Hemisphere, Chile, and observations collected at the Centro Astron\'omico Hispano Alem\'an (CAHA) at Calar Alto, operated jointly by the Max-Planck Institut f\"ur Astronomie and the Instituto de Astrof\'isica de Andaluc\'ia (CSIC).}}
   \subtitle{Indications for a young system with a forming brightest galaxy}

   \author{A. Nastasi
          \inst{\ref{MPE}}
          \and
	  R. Fassbender
          \inst{\ref{MPE}}
          \and
          H. B\"ohringer
          \inst{\ref{MPE}}
          \and
          R. \v{S}uhada
          \inst{\ref{MPE}}
          \and
          P. Rosati
          \inst{\ref{ESO}}
          \and
          D. Pierini
          \inst{\ref{guest}}
       	  \and
	  M. Verdugo
	  \inst{\ref{MPE}}
	  \and
          J.S. Santos
          \inst{\ref{ESAC}}
          \and
          A.D. Schwope
          \inst{\ref{AIP}}
          \and
          A. de Hoon
          \inst{\ref{AIP}}
           \and
          J. Kohnert
          \inst{\ref{AIP}}
          \and
          G. Lamer
          \inst{\ref{AIP}}
          \and
          M. M\"uhlegger
          \inst{\ref{MPE}}
          \and
          H. Quintana
          \inst{\ref{PUC}}
          }

  \institute{Max-Planck-Institut f\"ur extraterrestrische Physik (MPE),
              Giessenbachstrasse~1, 85748 Garching, Germany \\
              \email{alessandro.nastasi@mpe.mpg.de} \label{MPE}
            \and
	 Guest astronomer at the MPE \label{guest}
          \and
            Astrophysikalisches Institut Potsdam (AIP),
            An der Sternwarte~16, 14482 Potsdam, Germany \label{AIP}
        \and
         European Southern Observatory (ESO), Karl-Schwarzschild-Str.~2, 85748 Garching, Germany \label{ESO}
        \and
        Departamento de Astronom\'ia y Astrof\'isica, Pontificia Universidad
        Cat\'olica de Chile, Casilla 306, Santiago 22, Chile \label{PUC}
	\and
	European Space Astronomy Centre (ESAC)/ESA, Madrid, Spain \label{ESAC}
             }

   \date{Received ...; accepted ...}

 
  \abstract
   {We report on the discovery of a galaxy cluster at $z\!=\!1.490$ originally selected as an extended X-ray source in the \XMM Distant Cluster Project. Further observations carried out with the VLT-FORS2 spectrograph allowed the spectroscopic confirmation of seven secure cluster members, providing a median system redshift of $z = 1.490 \pm 0.009$. The color magnitude diagram of XMMU\,J0338.8+0021 reveals the presence of a well populated red sequence with z$-$H\,$\approx$\,3, albeit with an apparent significant scatter in color. Since we do not detect indications for strong star formation activity in any of the objects, the color spread could indicate different stellar ages of the member galaxies. In addition, we found the brightest cluster galaxy in a very active dynamical state, with an interacting, merging companion located at a physical projected distance of $d \approx 20$kpc. From the X-ray luminosity we estimate a cluster mass of $M_{200}\!\sim\! 1.2 \times 10^{14}\,\mathrm{M_{\sun}}$. The data seem to suggest a scenario in which XMMU\,J0338.8+0021 is a young system, possibly caught in a moment of active ongoing mass assembly.}


   \authorrunning{A. Nastasi et al.}
   \maketitle

\section{Introduction}
\label{par:intro}
\begin{figure*}[t]
\centering
\includegraphics[height=8.8cm, clip=true]{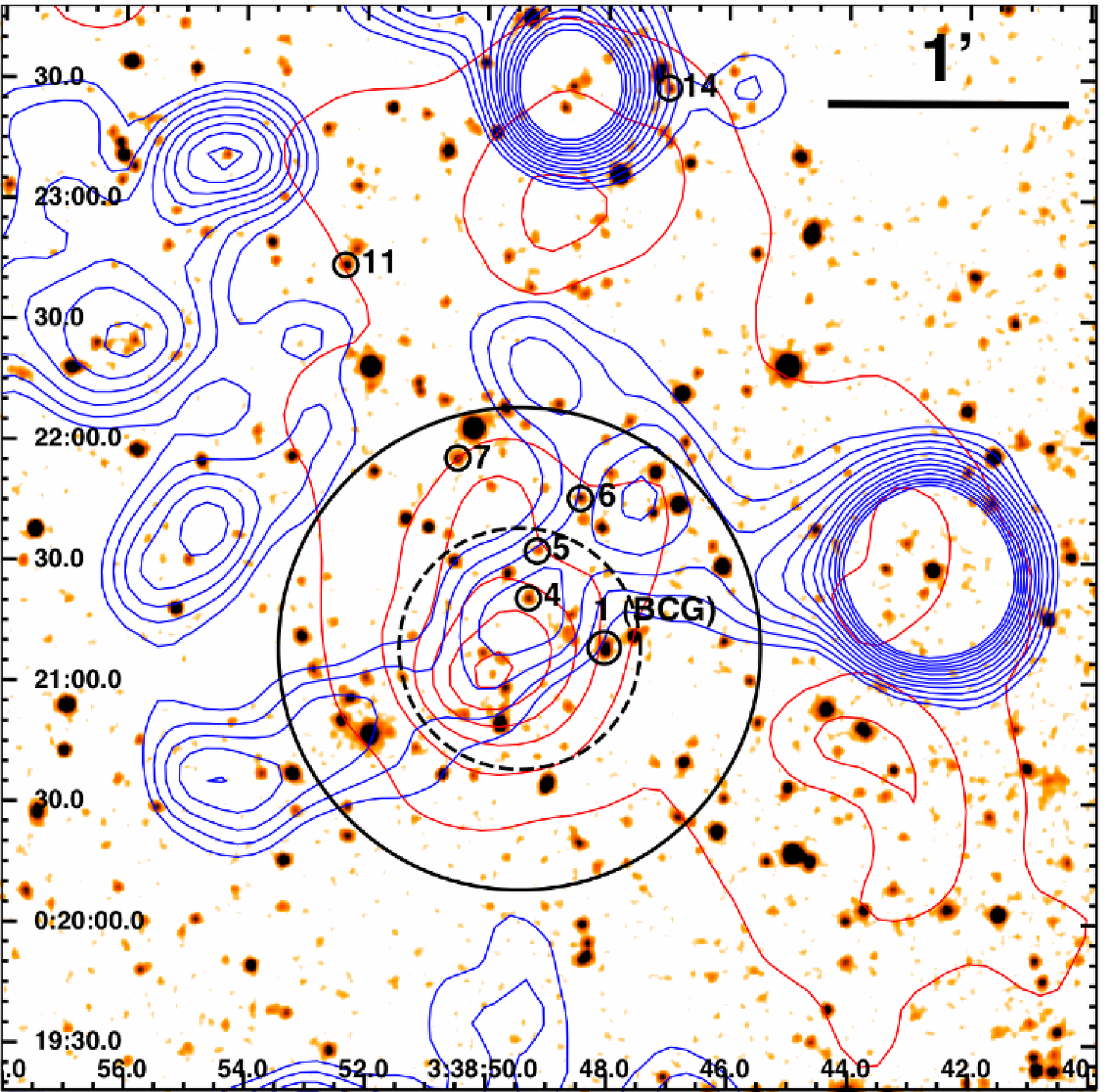}
\includegraphics[height=8.8cm, clip=true]{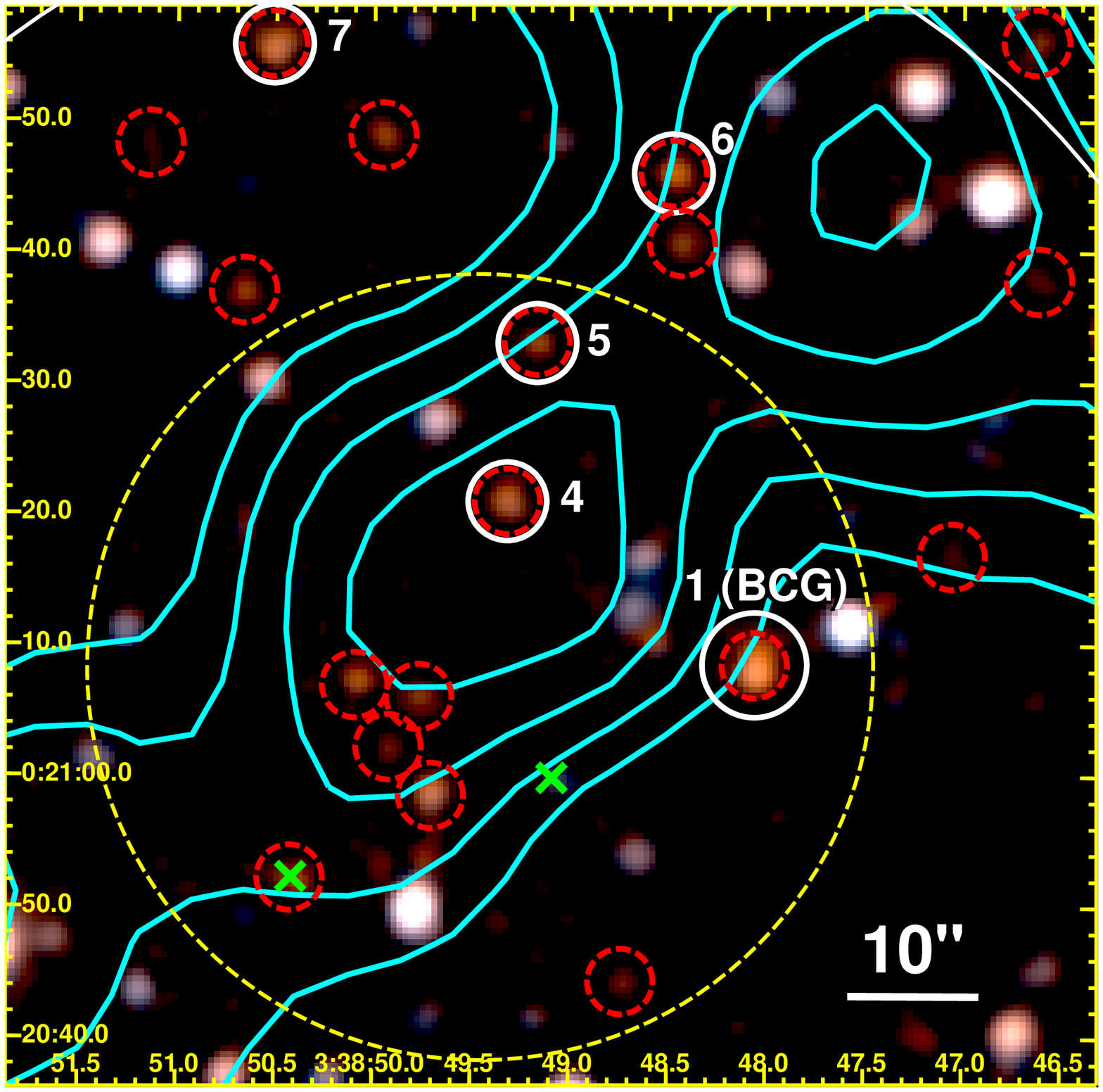}
\caption{\textit{Left:} H-band image (4.5\arcmin \ side length) of the environment of the galaxy cluster XMMU\,J0338.8+0021 at $z\!=\!1.490$. The \XMM detected X-ray emission is shown by the log-spaced blue contours, with the four lowest ones corresponding to significance levels of 2, 2.7, 3.6, 4.6\,$\sigma$ above the mean background. The associated overdensity of color-selected red galaxies is displayed by the red contours, showing linearly spaced significance levels of 2-20\,$\sigma$. The big, black circles indicate the 0.5\arcmin/1\arcmin \ radii around the X-ray center; the small black ones mark the spectroscopically confirmed cluster members, with the corresponding IDs. The image was smoothed with a 3 pixel (1.3\arcsec) Gaussian kernel. \textit{Right:} z+H-band color composite image of the core region of the cluster. X-ray contours are displayed in cyan, red galaxies with $2.6\!\le$z$-$H$\le\!3.5$ are marked by red dashed circles, spectroscopic members by white circles and interlopers (i.e. spectroscopically confirmed non-members) by green crosses. The 30\arcsec \ radius from the X-ray center is indicated by the yellow circle. The white ruler represents the beam size (FWHM) of \XMM at the observed off-axis angle of $\sim$5.6\arcmin.}
\label{fig:optical_view}
\end{figure*}
The search for distant ($z > 0.8$) galaxy clusters has experienced a remarkable boost recently. By means of various selection methods, several clusters have been found in the \textit{redshift desert}, at $z \geq 1.5$.
\cite{Papovich2010} and \cite{Tanaka2010} independently reported on a new IRAC selected cluster at $z=1.62$, characterized by a well populated and tight red sequence.

Further multi-band analysis revealed that the star formation activity in this cluster increases with the environment density, with star forming galaxies being mainly localized in the center of the system \citep{Tran2010}. An analogous trend has been observed in three other X-ray selected galaxy clusters, two newly confirmed at $z=1.56$ \citep{Fassbender2011} and $z=1.58$ \citep{Santos2011} and one at $z=1.46$ \citep[][]{Hilton2009, Hilton2010a}. For these systems, however, the galaxy red sequence is not fully established yet, revealing that the members are still building their stellar mass via star formation activity.
The situation is different in clusters found so far at $z < 1.4$, where active star forming galaxies preferentially reside in the cluster outskirts \citep[e.g.][]{Balogh2004c, Lidman2008, Patel2009}. However, it is expected that more distant galaxy clusters exhibit a larger number of star forming galaxies in their central regions \citep[e.g.][]{Hopkins2008g, Rettura2010}. In addition, early-type galaxies (ETG) in clusters are known to have completed the formation of their stellar populations much faster with respect to ETG in the field and the red sequence is expected to fade by $z\approx2$ \citep{Gobat2008}. The latter prediction is consistent with the recent results in \cite{Gobat2011}, where the discovery of the most distant cluster known to date, at $z=2.07$, is reported. In this system the population of passive galaxies shows a very large intrinsic scatter in the $Y-K_s$ colour, with an apparent absence of an established red sequence. This suggests that observations are now approaching the early stages of galaxy formation in the densest environments.

All these new findings highlight the importance to detect new distant galaxy cluster study targets for a systematic investigation of those processes which drive galaxy evolution in different environments.

In this Letter we report on the discovery of an X-ray selected, spectroscopically confirmed galaxy cluster at z=1.490, found within the \XMM Distant Cluster Project (XDCP).
We first present the X-ray data (\S \ref{par_s2:XMM_Xray}) followed by a discussion of the near-infrared (NIR) imaging observations (\S \ref{par_s2:imaging}) and the optical spectroscopy results (\S \ref{par:spectroscopy}). In Section \ref{par:assembling} we will discuss the dynamical state of the system and its galaxy population. Finally, a summary and conclusions are reported in \S \ref{par:conclusion}.

For this paper, we assume a concordance $\Lambda$CDM cosmology, with $H_0 =$ 70 km/s, $\Omega_\Lambda =$ 0.7, $\Omega_m =$ 0.3 and $w=-1$. For the given cluster redshift, the angular scale is $8.46$ kpc/\arcsec.
Magnitudes are given in the Vega system.
\section{Observations, data analysis, and results}
\label{par:obs_phot_X}
The cluster XMMU\,J0338.8+0021 (hereafter \79com) was discovered within the \XMM Distant Cluster Project (XDCP), a serendipitous archival X-ray survey focussed on the identification of X-ray luminous systems at $z\!>\!0.8$ \citep[e.g.][]{Mullis2005a,HxB2005a,Fassbender2007, Fassbender2011}.
\subsection{X-ray selection with \XMM}
\label{par_s2:XMM_Xray}
The X-ray source associated with the cluster \79com was initially observed in 2002, in the \XMM field with observation identification number (OBSID) 0036540101 targeting the quasar SDSS033829.31+002156 (z\,$=$\,5.02) with a nominal exposure time of 22.9\,ksec. It was detected as weak source with \texttt{SAS} v6.5 at an off-axis angle of 5.6\arcmin, with a source significance of about 5\,$\sigma$ and an extent significance of about 2.5\,$\sigma$.

We have reprocessed the field  with \texttt{SAS}\,v10.0 for a more accurate source characterization. After applying a strict two-step flare cleaning process for the removal of high background periods, a clean net exposure time of 16.4/15.9\,ksec remained for the two EMOS cameras and 7.8\,ksec for the PN instrument.
Figure\,\ref{fig:optical_view} shows log-spaced X-ray contours (blue/cyan) of the cluster environment with significance levels of 2-16\,$\sigma$ derived from the adaptively smoothed combined images.
For the flux measurement we applied the growth curve analysis (GCA) method of \citet{HxB2000a} in the soft 0.5-2\,keV band and measured $f_{X,500}\simeq(3.0 \pm 1.8)\times 10^{-15}$\,erg\,s$^{-1}$\,cm$^{-2}$ in a 40\arcsec \ ($\simeq\!R_{500} $) aperture. At the cluster redshift, this translates into a 0.5-2 keV X-ray luminosity of $L_{X,500}\simeq(4.1 \pm 2.4)\times 10^{43}$\,erg/s.
Owing to the faintness of the source at the limit of detectability, with $\sim$\,33 net counts within an aperture of 24\arcsec, the determination of additional structural or spectral parameters is currently not feasible and the extended nature of the source is tentative. However, based on the M$- L_X$ scaling relation in \citet{Pratt2009a}, with a modified redshift evolution (\cite{Fassbender2011}; Reichert et al., submitted), a first rough estimate of the expected temperature ($T^{est}_X$) and mass ($M_{500}$) can be obtained, with $T^{est}_X\!\sim\!2.5$\,keV and $M_{500}\!\sim\! 0.8 \times 10^{14}\,\mathrm{M_{\sun}}$, corresponding to a total mass estimate of $M_{200} \sim 1.2 \times 10^{14}{M_{\sun}} $ under the assumption of a NFW profile.
According to the above $T^{est}_X$ value, a bolometric luminosity $L^{\mathrm{bol}}_{X,500}\simeq(1.1 \pm 0.6)\times 10^{44}$\,erg/s is finally estimated. We cannot rule out that a non-negligible part of the X-ray flux is due to contaminating point sources and thus the reported luminosities have to be taken as upper limits. All the estimated parameters based on the X-ray flux are reported in Table\,\ref{Tab:Xray_props}.
\begin{table}[ht] 
\caption{Main properties of XMMU\,J0338.8+0021}
\centering
\label{Tab:Xray_props} 
\begin{tabular}{l l}  
\hline\hline
RA		   & 03:38:49.5		\\
DEC		   & +00:21:08.1	\\
z		   & 1.490$\pm$0.009	\\
\hline
$f_X (0.5-2 keV)$  & $(3.0 \pm 1.8)\times10^{-14} \mathrm{erg\,s^{-1}\,cm^{-2}}$\\
$L_X (0.5-2 keV)$  & $(4.1 \pm 2.4)\times10^{43} \mathrm{erg\,s^{-1}}$	 \\ 
$L^{bol}_X $  	   & $(1.1 \pm 0.6)\times10^{44} \mathrm{erg\,s^{-1}}$	 \\ 
$M_{X,200} $  	   & $\sim 1.2\times10^{14}\,\mathrm{M_{\sun}}$		 \\ 
$R_{200}$	   & $\sim$ 590 kpc					 \\
$T^{est}_X $  	   & $\sim$ 2.5 keV			 		 \\ 
\hline         
\end{tabular}
\end{table}
\subsection{Near-infrared follow-up imaging}
\label{par_s2:imaging}
\begin{figure}[h!]
\centering
\includegraphics[width=9.5cm]{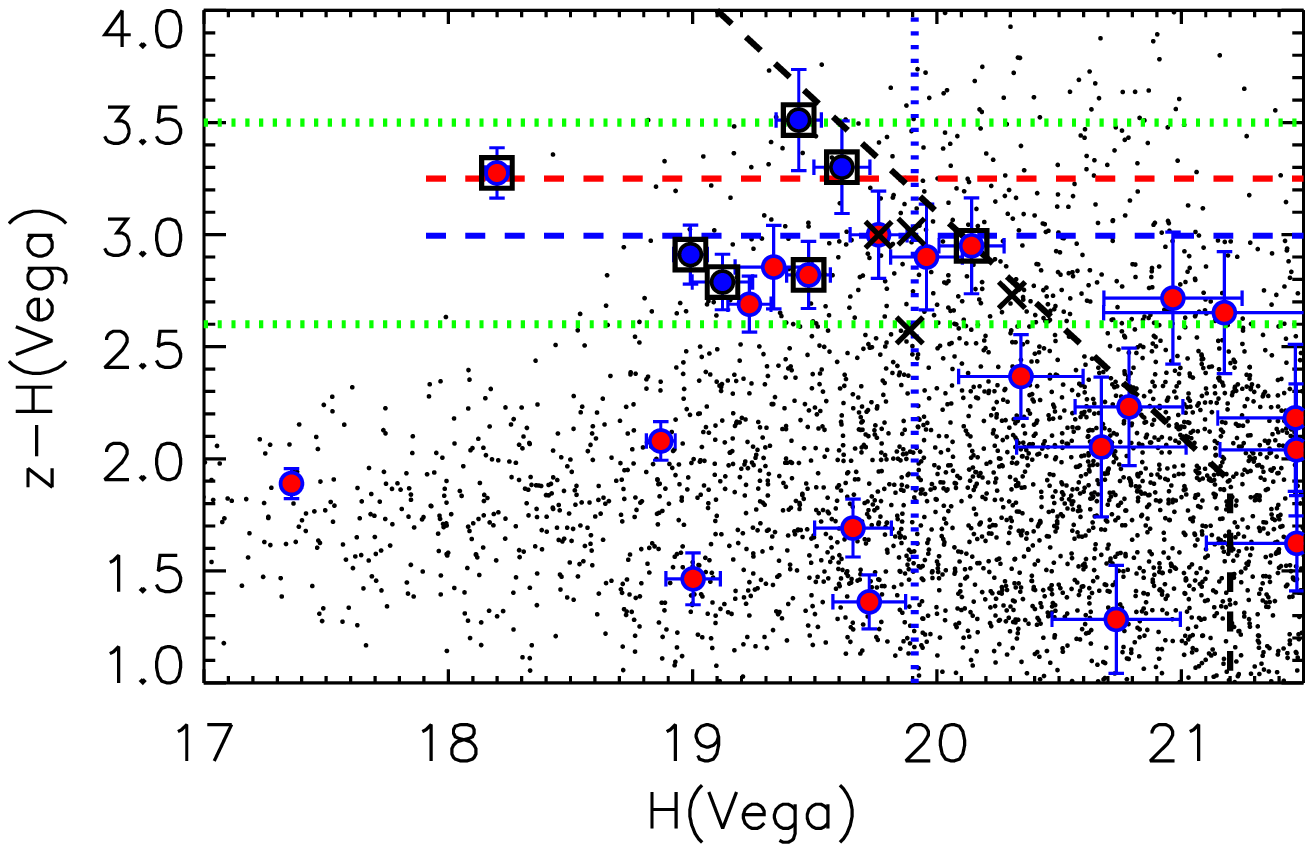}
\includegraphics[width=9.0cm]{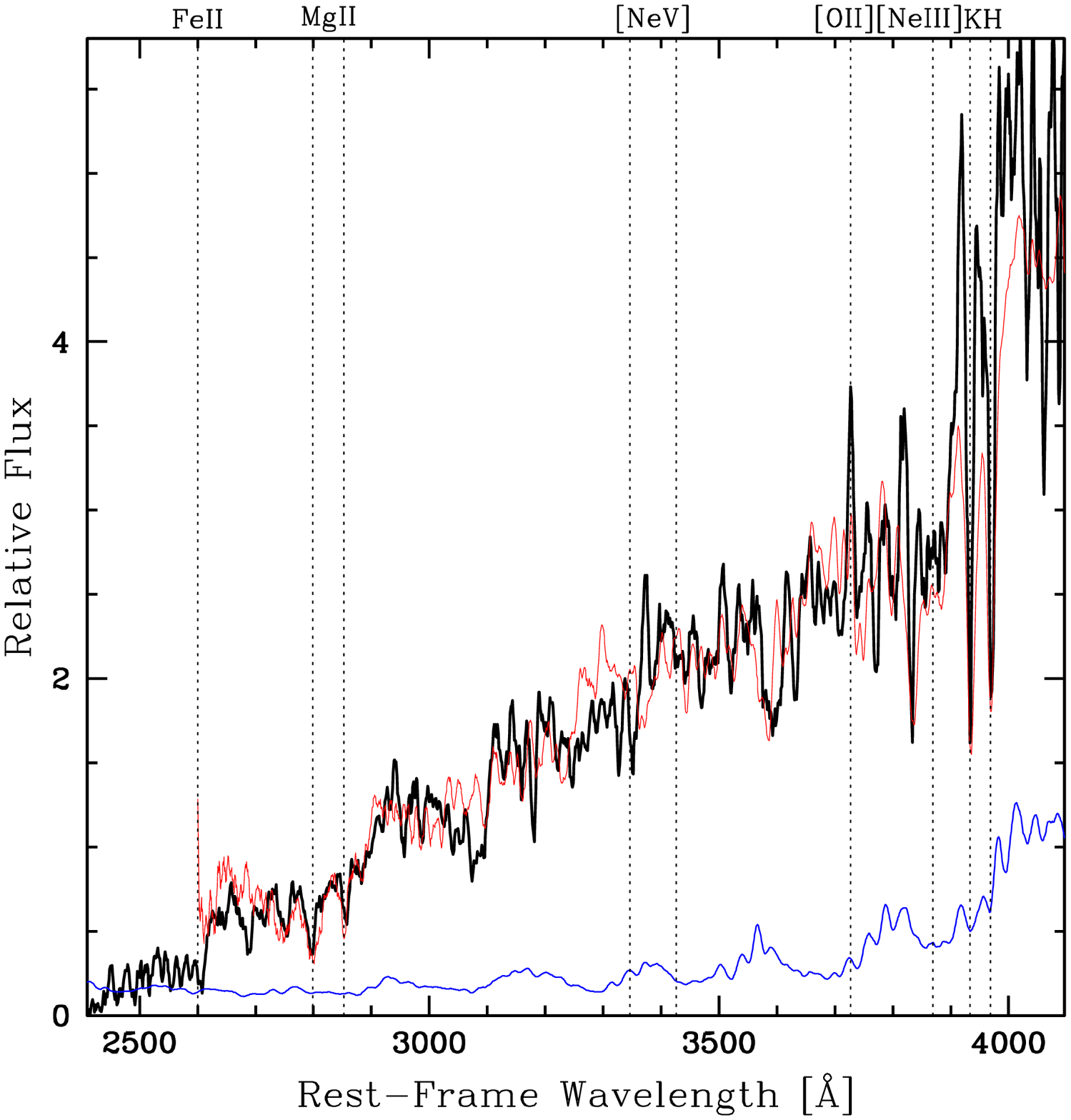}
\caption{\textit{Top}: Color-magnitude diagram of the cluster field. All galaxies with $D_{proj} \leq$\,30\arcsec \ are displayed in red. The seven secure spectroscopic cluster members are marked by black squares (with blue dots for $D_{proj} >$\,30\arcsec) and are all located within a color range 2.6\,$\leq$\,z$-$H\,$\leq$\,3.5 (green dotted lines), with an average color of z$-$H\,$\sim$\,3.
The interlopers found in the above color cut are marked by black crosses. Two simple stellar population (SSP) models for the cluster redshift are overplotted at z$-$H\,$\sim$\,3, with solar metallicity and formation redshift of $z_f\!=\!5$ (red dashed line) or $z_f\!=\!3$ (blue dashed). The vertical dotted blue line indicates the apparent magnitude of an L* galaxy at $z\!=\!1.49$, black dashed lines show the 50\%-completeness limit. \textit{Bottom}: Composite rest frame spectrum of the seven members galaxies. The main features are marked by dotted lines and the total noise spectrum is shown in blue. An LRG template based on $z \sim 0.5$ luminous red galaxies \citep{Eisenstein2003} is overplotted in red to show the good match with the low-z template. The spectrum is smoothed with a 7 pixel boxcar filter.}
\label{fig:CMD_StakedSpec}
\end{figure}

\79com was photometrically confirmed as a significant overdensity of very red galaxies coincident with the X-ray position using the wide-field (15.4\arcmin$\times$15.4\arcmin) near-infrared (NIR) camera OMEGA2000 at the Calar Alto 3.5m telescope \citep{Bailer2000a}.
We obtained medium-deep H-band  (50\,min) and z-band (53\,min) images of the \79com field on January 5, 2006 under good, but non-photometric, conditions with a seeing of 1.2\arcsec. Re-observations with the z-filter were performed (5\,min) on October 30, 2006 in photometric conditions for the target and designated SDSS standard stars \citep{Smith2002a} in order to allow a photometric calibration. The data were reduced with a designated OMEGA2000 NIR pipeline \citep{Fassbender2007}. 
The \texttt{Sextractor} \citep{Bertin1996a} photometry was calibrated to the Vega system using 2MASS point sources \citep{Cutri2003a} in H, and the SDSS standard star observations in z, cross-calibrated with SDSS photometry in the science field. 
The limiting (Vega) magnitudes (50\% completeness) were determined to be $H_\mathrm{lim}\!\sim\!21.2$ and $z_\mathrm{lim}\!\sim\!23.1$, corresponding to the expected apparent magnitudes of passively evolving galaxies at $z\!\sim\!1.5$ with absolute magnitude of $M$*+1.3 in H and $M$* in z.
Figure\,\ref{fig:optical_view} (left) shows an H-band image of the cluster field (4.5\arcmin$\times$4.5\arcmin), with overlaid X-ray contours (blue), density contours of color selected galaxies with $2.6\!\le$\,z$-$H\,$\le\!3.5$ (red), and marked spectroscopic cluster members (small black circles). A zoom onto the core region is shown in Fig.\,\ref{fig:optical_view}, right.
The z$-$H versus H color magnitude diagram (CMD) of the field is shown in Fig.\,\ref{fig:CMD_StakedSpec} (top).
\section{Spectroscopic analysis}
\label{par:spectroscopy}
\subsection{Data reduction}
\label{par_s2:spec_dataRed}
\begin{figure*}[ht]
\centering
\includegraphics[width=12cm]{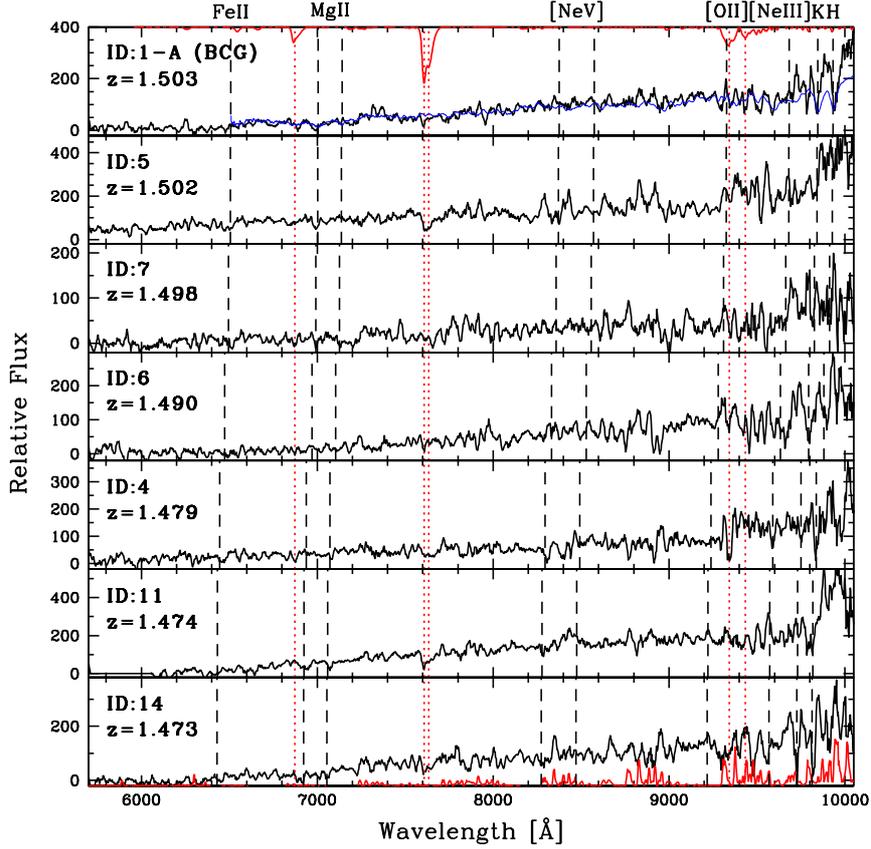}
\caption{1-D spectra of the seven confirmed members of XMMU\,J0338.8+0021, smoothed with a 7 pixel boxcar filter. For each spectrum, an identification label (ID) and the measured redshift is reported. For the brightest component of the BCG (ID:1-A) the same LRG spectrum template used in Fig.\ref{fig:CMD_StakedSpec} (bottom) is overlaid in blue for comparison. In red, the sky features in absorption (top) and emission (bottom) are shown. The position of the main spectral features, shifted according to the corresponding redshift values, are marked by the vertical dashed lines.}
\label{fig:7spectra}
\end{figure*}
\begin{table*}[ht]
\caption{Properties of spectroscopic members of XMMU\,J0338.8+0021}          
\centering  
\begin{tabular}{c c c c c c c c c l}   
\hline\hline            
ID  &      RA       &	   DEC	   &    z   & $\sigma_z$ & \multicolumn{2}{c}{Projected Distance} & \textit{H} & \textit{z$-$H} & Main spectral \\ 
    &	            &		   &        &            &(arcsec)      &(kpc)     & (Vega)     &    (Vega)    &   features    \\
\hline
 1-A&  03:38:48.048 & +00:21:08.06 & 1.5031 & 0.0006     & 20.8     	& 175.8    & 18.25      & 3.30         &  MgII, FeII, CaH/K \\
 5  &  03:38:49.159 & +00:21:32.77 & 1.5024 & 0.0005     & 24.9         & 210.5    & 20.19      & 2.97         &  MgII, FeII, CaK   \\
 7  &  03:38:50.483 & +00:21:55.65 & 1.4976 & 0.0007     & 49.8         & 421.4    & 19.17      & 2.81         &  MgII, FeII (weak), [OII] (weak)\\
 6  &  03:38:48.449 & +00:21:45.78 & 1.4899 & 0.0006     & 40.4         & 341.9    & 19.49      & 3.54         &  FeII, CaH/K      \\
 4  &  03:38:49.286 & +00:21:20.43 & 1.4788 & 0.0013     & 12.6         & 106.7    & 19.53      & 2.84         &  CaH/K            \\
 11 &  03:38:52.357 & +00:22:43.76 & 1.4736 & 0.0005     & 104.9        & 887.7    & 19.04      & 2.94         &  MgII, [OII] (weak), [NeIII], CaH/K\\
 14 &  03:38:46.998 & +00:23:27.98 & 1.4730 & 0.0007     & 144.5        & 1222.2   & 19.66      & 3.33         &  MgII, [OII] (weak), CaH/K\\
\hline \\ 
\label{Tab:properties} 
\end{tabular}
\end{table*}
\79com was observed with the FORS2-VLT spectrograph on December 2009 (program ID:84.A-0844) with seeing conditions varying between 0.7\arcsec and 1.1\arcsec. A single MXU slit-mask (field of view $6.8' \times 6.8'$) with a total of 42 1\arcsec\,width slits was used. The chosen instrument setup with the 300I grism without blocking filter provides a wavelength coverage of $5800-10500$\AA{}, with a resolution of $R=660$. The total net integration time was 8.4ks.
Raw data were reduced with a new FORS2 adaption of the \textit{VIMOS Interactive Pipeline Graphical Interface} (VIPGI, \cite{Scodeggio2005}), which includes all the standard reduction steps: bias subtraction, flat-field corrections, image stacking and wavelength calibration by means of a Helium-Argon reference spectrum (Nastasi et al., in prep.).

A total of 45 1-dimensional spectra were obtained, with a final calibration uncertainty of $\sim$1\AA{}.
\begin{figure*}[ht]
  \centering
  \includegraphics[height=8.5cm, clip=true]{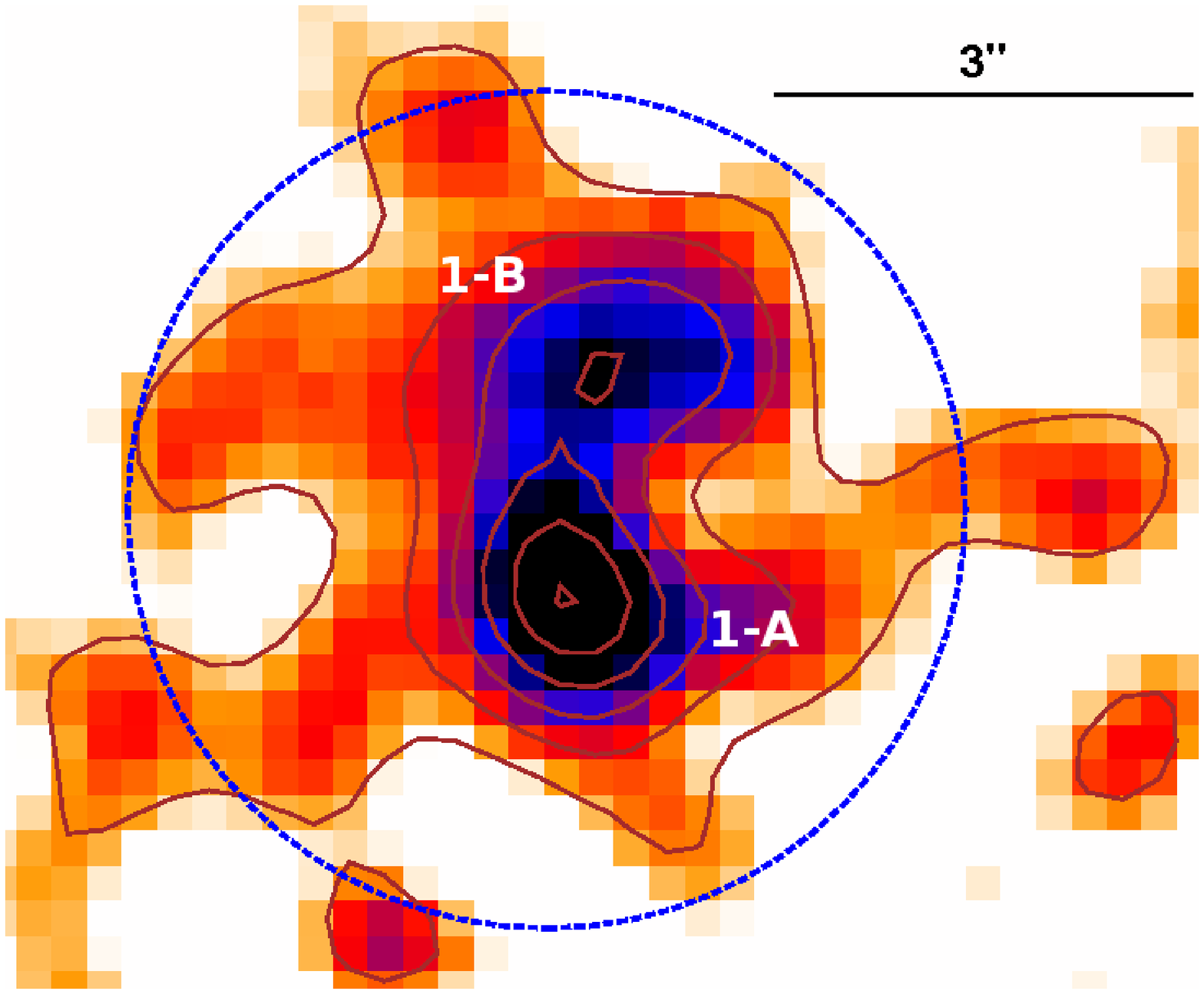}
  \includegraphics[height=8.5cm, clip=true]{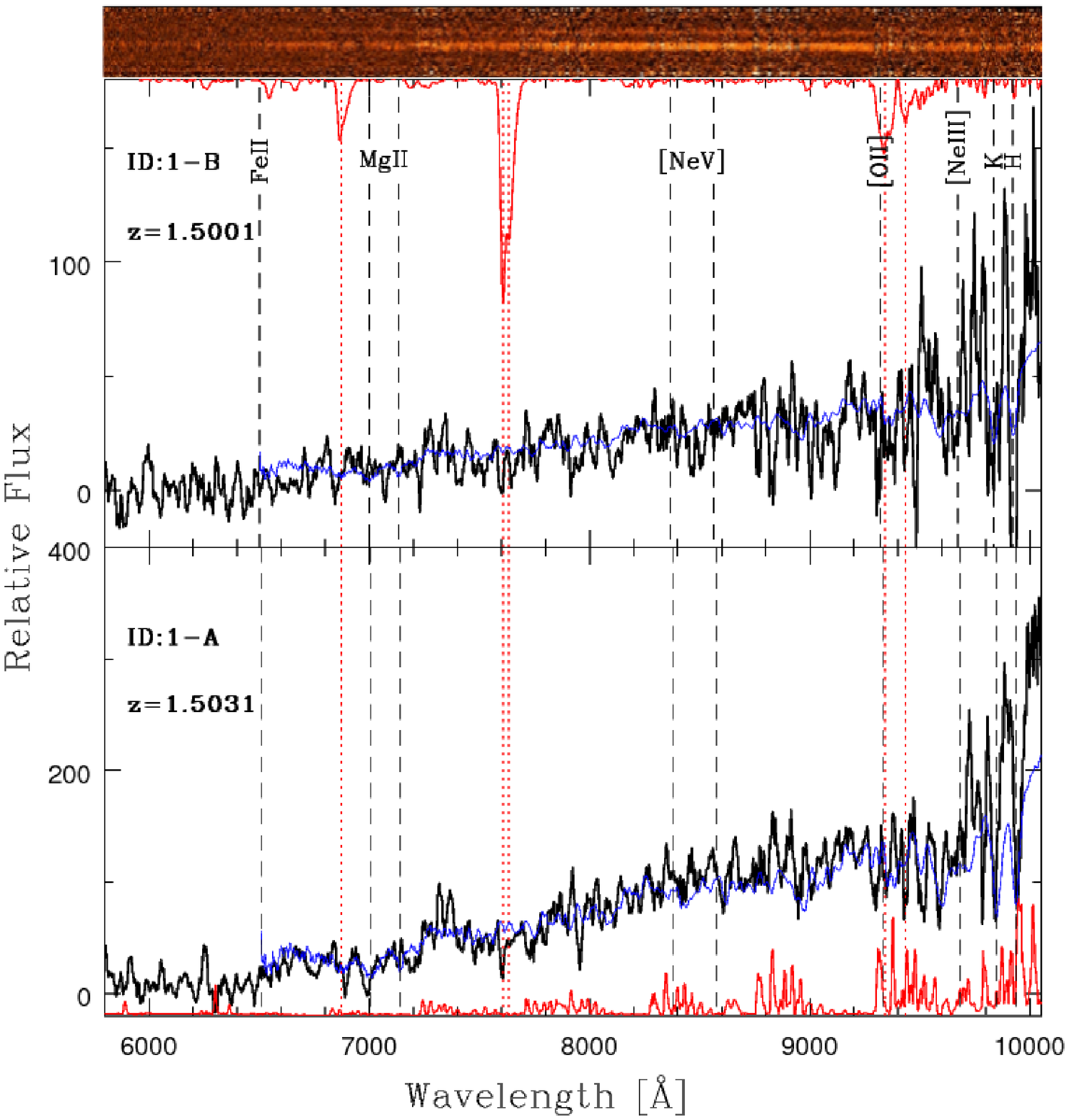}
  \caption{\textit{Left:} Close-up z-band image of the BCG of XMMU\,J0338.8+0021, smoothed with a 0.7 arcsec gaussian filter. The contours indicate linearly spaced isophotes of the image cutout and the blue circle refers to an area with 3\arcsec of radius. The presence of a double-core system is clearly visible, revealing the brightest galaxy is experiencing an ongoing assembling process. \textit{Right:} 2-D reduced spectra (\textit{top}) and the 1-D extracted ones (\textit{bottom}) for the two BCG components (ID: 1-A, 1-B). In the bottom panel, the ID number and the measured redshift for each spectrum is reported in the top left corner, and the corresponding best-fit LRG template is overlaid in blue. Because of the low signal-to-noise ratio obtained for ID:1-B (whose spectrum is barely distinguishable in the top panel), its redshift value is only tentative. Both spectra are smoothed with a 7 pixel boxcar filter.}
         \label{fig:BCG}
\end{figure*}
\subsection{Redshift measurements}
In order to assess the redshift for each extracted spectrum, the graphical VIPGI tools were initially used for a quick visual inspection. Then, the \texttt{EZ} software \citep{Garilli2010} and the \texttt{RVSAO/IRAF} package were also used with a set of different galaxy spectra templates.
The analysis confirmed the existence of seven galaxies with concordant redshifts in the range 1.473\,$\leq$\,z\,$\leq$\,1.503, six of which with a projected distance from the X-ray emission peak $D_{proj} < 2$\arcmin\, and rest frame velocity offsets of $| \Delta v | < $\,2000 km/s with respect to the median redshift. For the BCG, we extracted the spectra of both components ID:1-A and ID:1-B (Fig.\,\ref{fig:BCG}, right), distinguishable in the 0.86\arcsec \,seeing z-band preimage of Fig.\,\ref{fig:BCG} (left). For the second component (ID:1-B) only a tentative redshift measurement of $z = 1.5001$ can be obtained owing to its low signal-to-noise ratio.
The extracted spectra and the main properties of the seven confirmed members are shown in Fig.\,\ref{fig:7spectra} and reported in Table\,\ref{Tab:properties}, respectively. Several spectral features are identifiable, albeit strong emission and absorption sky lines affect the quality of the individual spectra.
This allows us to confirm the presence of a bound system of, at least, seven galaxies with a median redshift of $z = 1.490 \pm 0.009$. Using the biweight estimator \citep{Beers1990a} we obtain a crude velocity dispersion estimate of $\sigma_v \approx 670$ km/s in the cluster rest frame, albeit with an associated large uncertainty owing to the poor statistics.

In order to enhance the faintest features, an average, noise-weighted composite spectrum of the seven members was created. The result is shown in Fig.\,\ref{fig:CMD_StakedSpec} (bottom) and compared with an LRG template spectrum (red), derived for 726 luminous red galaxies at $0.47 < z < 0.55$ by \cite{Eisenstein2003}.
\section{A young cluster in formation?}
\label{par:assembling}
From the CMD in Fig.\,\ref{fig:CMD_StakedSpec} (top) we note that the red sequence of \79com appears well populated. The 11 red galaxies, with colors 2.6\,$\leq$\,z$-$H\,$\leq$\,3.5 and brighter than the completeness limit, have a median (average) color of z$-$H$=$2.9\,mag (3.0\,mag), which is consistent with the model prediction of z$-$H\,$\simeq$\,3.0 mag for a stellar formation redshift of $z_f =$\,3, which implies an evolved red sequence population, as seen in other distant clusters (XMMUJ1229, \cite{Santos2009a}; XMMUJ2235, \cite{Strazzullo2010}). From the observed color spread of $\delta _{z-H} ^{obs}$\,$\approx$\,0.25 mag, with an assumed flat slope and a Monte-Carlo estimated contribution of the photometric errors of (0.16$\pm$0.04) mag, we obtain a first order estimate of the intrinsic red sequence scatter of $\delta ^{int} _{z-H}\sim$\,(0.19$\pm$0.04) mag. This is significantly larger than in the above cases, e.g. 0.08 for XMMUJ2235, but consistent with the findings of \cite{Hilton2009} for their z=1.46 cluster. The estimate for the above intrinsic scatter does not change if a non-flat slope (e.g. the one of XMMUJ2235 given by \cite{Strazzullo2010}) for the fit is used. 
This larger scatter can be explained with stellar formation age differences of the bright end of the red galaxy population of the order of 0.5\,-\,1\,Gyr, as discussed in \cite{Hilton2009}.

The seven spectroscopic members show weak or no signs of ongoing star formation activity, with a measured equivalent width of the [OII] line in their spectra of $|EW_{[OII]}| \le$\,5\AA{}. Assuming the above spectral feature as a proxy of their star formation rate\footnote{The presence of the [NeIII]$\lambda$3870 line for the ID:11 galaxy, however, suggests the presence of an obscured AGN for it.}, the measured values for $EW_{[OII]}$ indicate a very moderate intensity for such activity. The existence of a real, albeit weak, [OII] emission activity in the \79com members is confirmed by the presence of such a spectral feature in the composite spectrum of Fig.\ref{fig:CMD_StakedSpec}, bottom. 

On the red sequence we find a magnitude gap between the BCG and the second ranked member of $\Delta m_{12} \approx$\,0.8\,mag. From the study of \cite{Smith2010} in the local universe, this points towards a dynamically evolved cluster, while the apparent offset of the BCG with respect to the center of X-ray emission ($D ^{BCG} _{proj} \approx$\,175\,kpc) is a typical sign of a dynamically young clusters \citep{Haarsma2010}.
This is in line with the fact that the cluster shows indications of asymmetric X-ray emission with a low central surface brightness. A close-up look at the BCG, shown in Fig.\,\ref{fig:BCG} (left), shows a double-object morphology, indicating an active mass assembly process via a merger event (implying that the relatively large magnitude gap is a very recent feature). Because of the lack of detectable emission lines in ID:1-A and 1-B spectra, we can state that this coalescence is likely a dry merger. The overall color and magnitude of the BCG are consistent with the ones found in red, passive, brightest galaxies residing in low-redshift clusters. The case of \79com seems to suggest that we may be finally witnessing the active mass assembly epoch of BCGs, in contrast to the evolved counterparts observed at $z \la 1.4$ \citep[e.g.][]{Whiley2008, Collins2009}. Our observations support the hierarchical assembly scenario of BCGs, albeit at significantly earlier epoch than predicted from the cosmological simulation results of \citet{DeLucia2007a}.

The large scatter of the red sequence, the X-ray morphology and the spatial distribution of color-selected red galaxies point towards a younger cluster compared to e.g. XMMUJ2235, a scenario that will be critically tested by planned deeper observations.
\section{Summary and conclusion}
\label{par:conclusion}
We summarize our results as follows: 
\begin{itemize}
\item[$\bullet$] We reported on the study of a newly discovered galaxy cluster XMMU\,J0338.8+0021 at $z=1.490$. This system was selected as an extended X-ray source serendipitously detected within the XDCP.
\item[$\bullet$] By means of spectroscopic follow-up we can confirm seven cluster members, six of which with $D_{proj} < 1$\,Mpc and a rest frame velocity offset of $|\Delta v| < $\,2000\,km/s from the median redshift.
\item[$\bullet$] The H+z band data reveal the presence of a well populated, albeit quite spread, red sequence in the CMD, with a median color of z$-$H\,=\,2.9 mag and an intrinsic color scatter of $\delta _{z-H} ^{intr} \sim$\,0.19. Comparisons with SSP models suggest that the stellar populations of red sequence galaxies span the formation epoches in the redshift range $3 \la z_f \la 5$.
\item[$\bullet$] A BCG is clearly identifiable on the red sequence and found in an active mass assembly phase, likely via a dry merging process.
\item[$\bullet$] Our data overall suggest that \79com is a young galaxy cluster not yet fully evolved, as the faint X-ray emission appears to be morphologically disturbed and the large area over which its galaxies are found possibly indicates the presence of infalling structures.
\end{itemize}

Currently, the present evidence for \79com suggests partly an evolved cluster scenario and at the same time signs of a dynamically young system. More detailed studies are needed for a more complete characterization of this intriguing system and its components.

\begin{acknowledgements}
The \XMM project is an ESA Science Mission with instruments and contributions directly funded by ESA Member States and the USA (NASA).
The XMM-Newton project is supported by the Bundesministerium f\"ur Wirtschaft und Technologie/Deutsches Zentrum f\"ur Luft- und Raumfahrt (BMWI/DLR, FKZ 50 OX 0001), the Max-Planck Society and the Heidenhain-Stiftung. 
This research has made use of the NASA/IPAC Extragalactic Database (NED) which is operated by the Jet Propulsion Laboratory, California Institute of Technology, under contract with the National Aeronautics and Space Administration. 
This work was supported by the Munich Excellence Cluster Origin and Structure of the Universe (www.universe-cluster.de), by the DFG under grants Schw536/24-1, Schw 536/24-2, BO 702/16-1,16-2,16-3, and the German DLR under grant 50 QR 0802.
We thank the anonymous referee for its comments, which helped to improve this paper. We also thank Gabriel Pratt for its useful comments.
AN would like also to thank Angela Bongiorno and Michele Cappetta for fruitful discussions and helpful comments.
\end{acknowledgements}
\bibliographystyle{aa} 
\bibliography{Nastasi_biblio.bib}
\end{document}